\documentclass[aps,prl,twocolumn,superscriptaddress,reprint,showpacs]{revtex4-2}
\usepackage{graphicx}
\usepackage{setspace}
\usepackage{amsmath}
\usepackage{amssymb}
\usepackage{color}
\usepackage{array}
\usepackage{subfigure}
\usepackage{hyperref}
\usepackage{float}
\usepackage{lipsum}

\usepackage[all]{xy}
\newcommand{\RN}[1]{%
  \textup{\uppercase\expandafter{\romannumeral#1}}%
}

\begin{document}
\title{Heterodyne detection of low-frequency fields via Rydberg EIT with phase demodulation}

\author{Shenchao Jin}
\thanks{Shenchao Jin and Xiayang Fan contributed equally to this work.}
\author{Xiayang Fan}
\thanks{Shenchao Jin and Xiayang Fan contributed equally to this work.}
\author{Xin Wang}
\author{Yi Song}
\author{Yuan Sun}
\email[email: ]{yuansun@siom.ac.cn}
\affiliation{Shanghai Institute of Optics and Fine Mechanics, Chinese Academy of Sciences, Shanghai 201800, China}
\affiliation{University of Chinese Academy of Sciences, Beijing 100049, China}

\begin{abstract}
Recently, the rapid progress of quantum sensing research reveals that the Rydberg atoms have great potentials in becoming high-precision centimeter-scale antenna of low-frequency fields. In order to facilitate efficient and reliable detection of low-frequency fields via Rydberg atoms, we design, implement and analyze a special but low-cost and scalable method based on heterodyning processes under the condition of electromagnetically induced transparency (EIT) embedded in typical two-photon ground-Rydberg transition. Instead of relying on observing changes in absorption of light by Rydberg atoms, our method focuses on the phase modulation effect on the probe laser induced by the low-frequency fields via the Rydberg EIT mechanism and utilizes a demodulation process to accurately retrieve the signal. The general principles of our method apply to both electric and magnetic fields and it is even possible to realize the combination of both functionalities in the same apparatus. In particular, we experimentally demonstrate the full cycle of operations with respect to both cases. In the measurement of low-frequency electric fields, we discover that the Rydberg dipole-dipole interaction among atoms induce linear superposition of Rydberg states with different angular momentum that generates a first-order response corresponding to the signature of linear Stark effect. As the Rydberg atoms have excellent coupling strengths with electric fields, our results indicate that our method can hopefully reach high-precision performance for practical tasks in the future. 
\end{abstract}
\maketitle



In recent years, Rydberg atoms have been making solid progress in quantum sensing \cite{RevModPhys.89.035002, Holloway2024Review, Zhang2024Review} and quantum computation \cite{RevModPhys.82.2313}. The key advantage of natural or artificial atoms' Rydberg states is their much greater coupling strength with electric fields compared to ground states, yielding significant potential for high-precision, compact quantum sensors of electric signals. Quantum sensing of microwave \cite{Shaffer2012NP, PhysRevLett.111.063001, PhysRevApplied.5.034003, PhysRevLett.120.093201, PhysRevLett.121.110502, ZhangLinjie2020NP, Shi2023APL, Ding2023NC, Yanhui2024SA} and THz \cite{Weatherill2023Review} electric fields via Rydberg atoms has already become a standard routine in various applications \cite{PhysRevApplied.19.044079, Cox2023APL} and the pursuit of higher-level optic-electric integration has begun \cite{PhysRevApplied.15.014047, PhysRevApplied.15.014053, Fu10670032, Fu10746562}. Recently, it has been discovered that Rydberg atoms have the potential to serve as an excellent candidate for centimeter-scale antenna of low-frequency fields when equipped with the advanced quantum coherent control methods \cite{PhysRevApplied.13.054034}. Further development faces several major obstacles, including the shielding effect from alkali atoms coating the vapor cell walls and the scalability challenge in building larger systems to achieve the theoretically predicted high sensitivity \cite{Lim2023APL}. At this moment, many investigations in this direction rely on the built-in electrodes within the vapor cell to induce interaction between low-frequency electric fields and Rydberg atoms \cite{Zhaojianming2023OE}. 

On the other hand, quantum sensing of low-frequency and radio-frequency magnetic fields by optical magnetometry with alkali atom vapor has achieved superior performances \cite{PhysRevLett.95.063004, Pustelny2018297} over many years of efforts. The atomic magnetometer \cite{Budker2007review} achieves high sensitivity through atomic ensembles’ collective response to the magnetic field, with the prominent example of optically pumped magnetometer (OPM) \cite{PhysRevLett.89.130801, Zhai2023review} achieving sub-fT sensitivity in the spin-exchange relaxation-free (SERF) regime \cite{Romalis2003Nature, Romalis2010APL}. In OPMs, detecting low-frequency magnetic signals requires a relatively weak bias magnetic field, necessitating magnetic shielding and feedback control. An optimized bias field enables precise detection of $\sim$kHz signals, making it useful for practical tasks such as electromagnetic induction imaging \cite{Renzoni2021APL, Renzoni2023APL}. Alternatively, magnetometers based on nonlinear magneto-optical rotation (NMOR) \cite{PhysRevA.62.043403}, which share physical principles with coherent population trapping (CPT) \cite{RevModPhys.77.633}, typically operate below 100 Hz \cite{Xiao2020NC, Xiao2021APL} due to CPT’s narrow linewidth, while modifying parameters can extend the effective range to the kHz regime at the expense of sensitivity.

Aiming for a systematic, low-cost and scalable solution for high-precision detection of low-frequency electric field with Rydberg atoms, we propose, analyze and experimentally demonstrate a special method with heterodyne processes to resolve the signal, which works under the condition of electromagnetically induced transparency (EIT) embedded in typical two-photon ground-Rydberg transition. Our method functions through the phase modulation effect on the probe light induced by the low-frequency fields via the Rydberg EIT mechanism and utilizes a demodulation process to accurately retrieve the signal. The basic mechanisms work for the detection of both the magnetic and electric signals. The rest of paper is organized as follows. At first, we introduce the basic mechanisms of the new method. Then we demonstrate the proof-of-principle operation for the low-frequency magnetic signals with also example of response to frequency-modulated signal, and we adhere to the applying a constant steady bias field. Next, we present the experimental results of detecting low-frequency electric field by extending the same principle. Interestingly, a magnetic bias field can also help in this case. We will also include discussions about the general properties of Rydberg states of alkali atoms interacting with low-frequency electric signals. The sensitivity of measurement will be evaluated according to the well-established standard criteria of quantum sensing with Rydberg atoms.

\begin{figure}[h]
\centering
\includegraphics[width=0.46\textwidth]{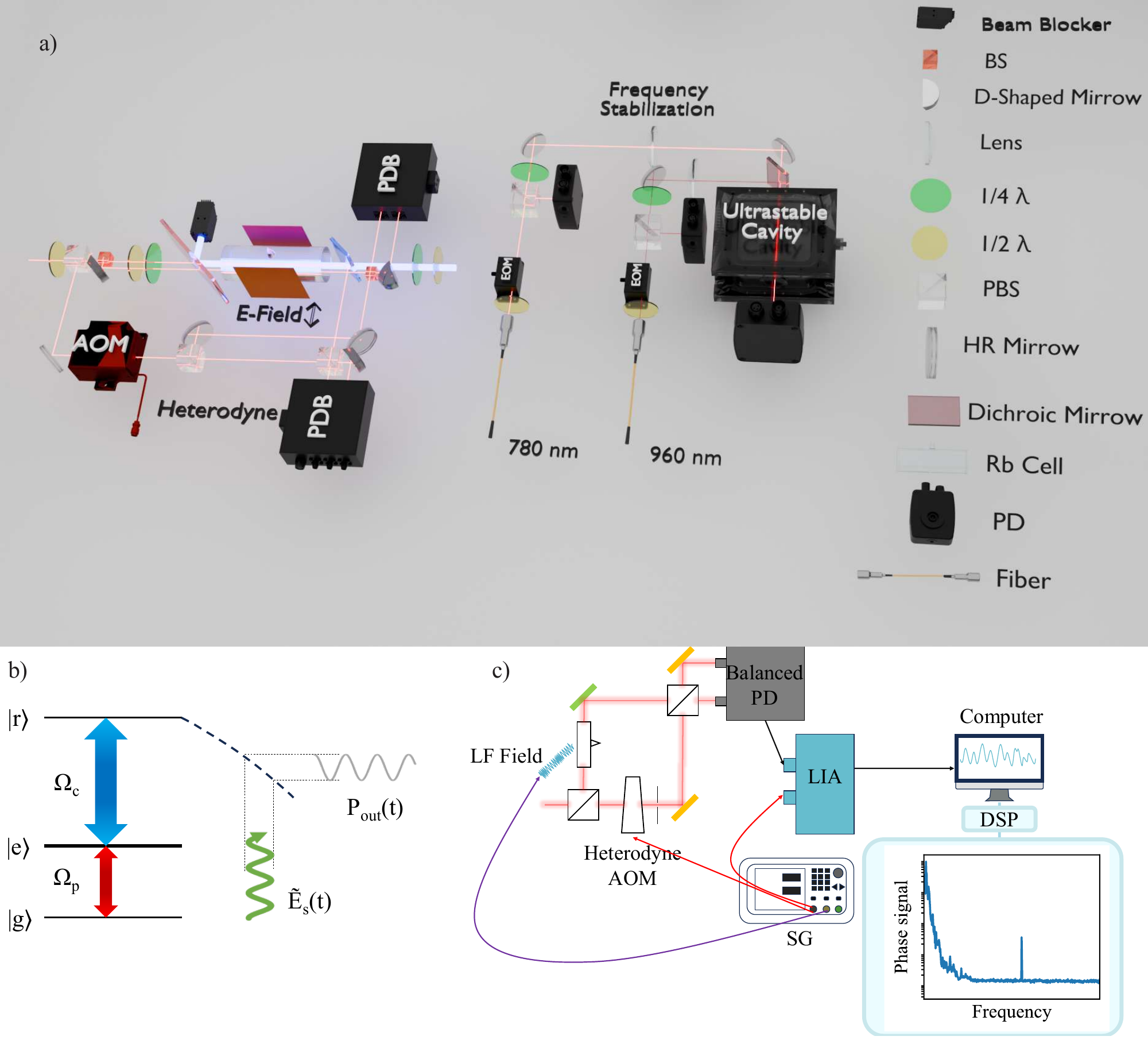}
\caption{(Color online) (a) Schematics of experimental setup. (b) Level diagram of idealized three-level two-photon ground-Rydberg transition. (c) Heterodyne, detection and post-processing parts of experiment apparatus, showing the basic ideas of relevant demodulation processes. AOM: acousto-optic modulator; SG: signal generator; PD: photodiode; LIA: lock-in amplifier; DSP: digital signal processor.}
\label{fig:basic_layout}
\end{figure}

We explain the basic principles via a typical three-level system with two-photon ground-Rydberg transition as shown in Fig. \ref{fig:basic_layout}(b). The probe laser always stays resonant with the transition $|g\rangle\leftrightarrow|e\rangle$ while the coupling laser of Rabi frequency $\Omega_\text{c}$ stays resonant with the transition $|e\rangle\leftrightarrow|r\rangle$. When the external field perturbs the Rydberg state $|r\rangle$, the coupling laser will subsequently experience a non-vanishing two-photon detuning $\Delta$. We can define $\alpha_0$ as unsaturated resonant absorption coefficient: $\alpha_0 = \frac{2\pi\nu_p N \mu_{ge}^2}{2\hbar c \epsilon_0\gamma_e}$ with $\nu_p$ as the probe laser frequency, $N$ as the atom number density, $\mu_{ge}$ as the atomic dipole moment of transition $|g\rangle\leftrightarrow|e\rangle$ and $\gamma_e$ as the decay rate of $|e\rangle$. The linear susceptibility $\chi$ contains the real part $\Re\chi$ and imaginary part $\Im\chi$ \cite{RevModPhys.77.633, RevModPhys.82.1041, Finkelstein_2023review}. Here we are interested in $\Re\chi(\Delta)$ which generates an effective phase change on the probe laser:
\begin{equation}
\label{eq:Re_chi}
\Re\chi(\Delta) = \frac{c\Delta|\Omega_\text{c}|^2 \alpha_0}{\pi\nu_p\gamma_e(\gamma_\text{eff}^2 + \Delta^2)} , \textit{ with }
\gamma_\text{eff} = \gamma_r + \frac{|\Omega_\text{c}|^2}{\gamma_e},
\end{equation}
where $\gamma_r$ is the decay rate of $|r\rangle$.

We prefer that the experiment operates under the strong coupling regime $|\Omega_\text{c}|^2\gg \gamma_e\gamma_r$ to yield a good transparent condition. Meanwhile, the performance of phase detection works well under the condition that the frequency variation range of the incoming signal stays within the transparency window of Rydberg EIT, which suits the case of low-frequency fields. The magnetic field perturbs the Rydberg state via Zeeman effect while the electric field perturbs the Rydberg state via Stark effect. Consequently, the low-frequency field drives $\Delta$ as in Eq. \eqref{eq:Re_chi} into a time-dependent term, effectively like the instantaneous response. For example, an incoming signal of frequency $\omega_\text{l}$ induces a phase modulation $\beta_\text{l}\cos(\omega_\text{l}t + \phi_\text{l})$ in the probe laser to the first order, where $\beta_\text{l}$ describes the strength of modulation and $\phi_\text{l}$ represents the initial phase. According to Eq. \eqref{eq:Re_chi}, the first-order effect when $|\Delta|$ is relatively small is given by:
\begin{equation}
\label{eq:EIT_1st_order}
\Re\chi(\Delta) = \frac{c|\Omega_\text{c}|^2 \alpha_0}{\pi\nu_p\gamma_e\gamma_\text{eff}^2}\Delta
+ \mathcal{O}(\Delta^3).
\end{equation}
On the other hand, higher order effects also exist as we have observed in experiments.

Fig. \ref{fig:basic_layout}(a) outlines the experimental setup. We choose $^{87}$Rb vapor under room temperature and the optical depth (OD) is about 0.2. The coherent ground-Rydberg excitation as well as the phase modulation and demodulation in the experiment all stem from a 480 nm laser and a 780 nm laser. The 780 nm laser couples the ground state $\left| g \right\rangle$ of 5$^2$S$_{1/2}$, $F=2$ and the excited state $\left| e \right\rangle$ of 5$^2$P$_{3/2}$, $F=3$ with beam waist of 0.3 mm. The 480 nm laser couples the excite state of $\left| e \right\rangle$ and the Rydberg state $\left| r \right\rangle$ with beam waist of 0.4 mm. In the magnetic field measurement, the vapor cell is of cylinder shape with 25 mm in diameter and 100 mm in length. In the electric field measurement, the vapor cell is of cuboid shape with $25~\mathrm{mm}\times 25~\mathrm{mm}\times 50~\mathrm{mm}$.

The frequency stabilization plays an important role in ensuring the performance, and both lasers are locked to one ULE high-finesse cavity made by Qingyuan Tianzhiheng Quantum Technology Co.,LTD. The 780 nm laser is at first split into the reference and probe beams. In the atom-laser interaction section, we divide the 780 nm probe beam into EIT probe and two-level-atom (TLA) probe beams, which are identical and parallel. The 480 nm laser counter-propagates with the EIT probe laser and they form the desired Rydberg EIT interaction. The TLA probe laser passes through the $^{87}$Rb atoms directly whose signal serves as the background. In the heterodyne section, the 780 nm reference receives 50 MHz frequency shift $\omega_\text{h}$ and subsequently mixes with the EIT probe laser. Then the beat signal contains the information of both the carrier $\omega_\text{h}$ and the incoming low-frequency signal: $\cos\big(\omega_\text{h}t + \beta_\text{l}\cos(\omega_\text{l}t + \phi_\text{l})\big)$. By two layers of demodulation efforts or lock-in amplification, the information of incoming low-frequency field can be recovered. In the first layer of demodulation, a Z\"urich lock-in amplifier extracts the effective low-frequency modulation component out from the beat signal oscillating around carrier frequency $\omega_\text{h}$. The second layer of demodulation focuses on the low-frequency range of interest, which generates the eventual output data via a second lock-in amplifier, spectrum analyzer or digital signal processor with similar purposes. As a comparison, the homodyne type method usually requires the careful construction and lock of an optical interferometer such as in the case of sensing microwave signals \cite{PhysRevApplied.19.064021}. In comparison, the heterodyne method for low-frequency detection avoids the trouble of setting up and locking interferometer while bearing more robustness in practical implementation.

\begin{figure}[h]
\centering
\includegraphics[width=0.42\textwidth]{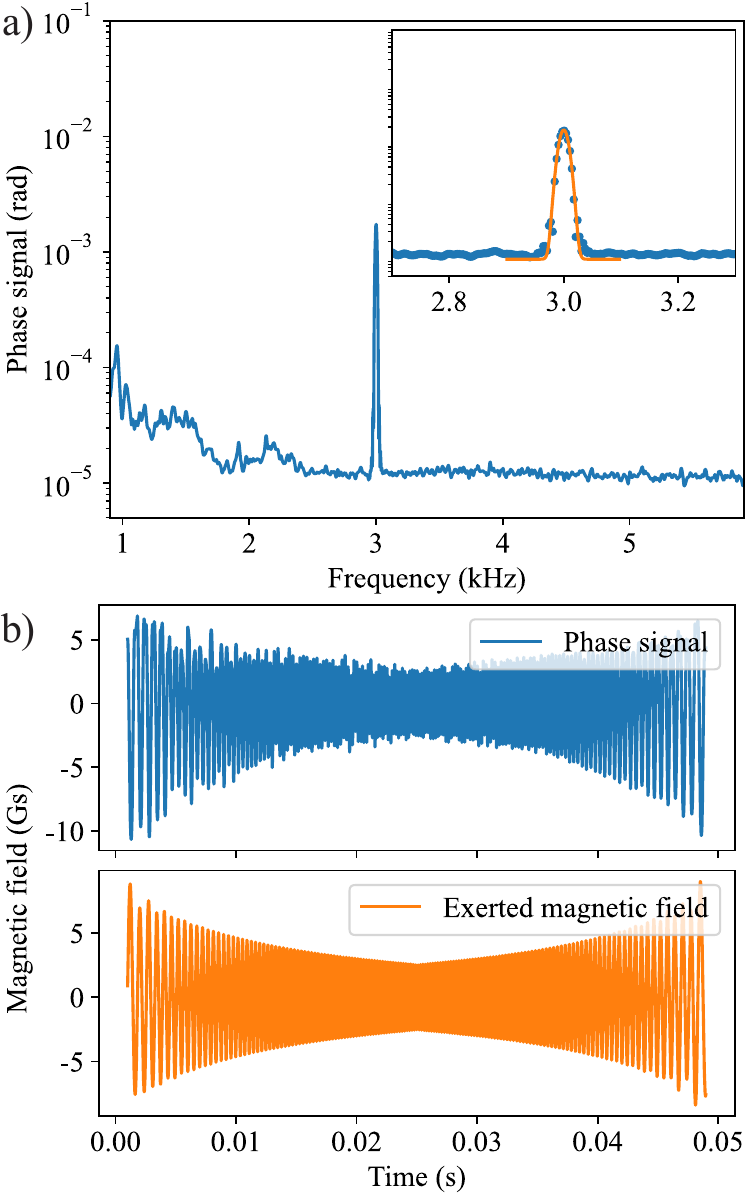}
\caption{(Color online) Results of low-frequency magnetic field. (a) Spectrum of demodulated EIT probe phase signal. Inset shows the Lorentz fitting. Here the signal-to-noise ratio (SNR) is 150.5, and the estimated sensitivity at 3.0 kHz is $0.87~ \mathrm{\mu T/\sqrt{Hz}}$.
(b) Performance with frequency-modulated (FM) signal, with comparison of the recovered signal (up) and the exerted magnetic field (down). The linear-chirped signal has center frequency of 3.0 kHz, FM range of $\pm 2.0$ kHz and FM speed of 20.0 Hz. The signal undergoes a band-pass filter with 700 Hz - 6000 Hz. The effective sensitivity here is $3.73~ \mathrm{\mu T/\sqrt{Hz}}$.}
\label{fig:low_frequency_magnetic}
\end{figure}

Applying a dc bias field is a common approach, particularly as Rydberg atoms have been employed to measure dc electric fields \cite{PhysRevLett.82.1831, PhysRevA.84.033402} since early times and dc bias electric field has already assisted microwave detection \cite{Shi2023APL}. Nevertheless, instead of using dc electric field we choose a set of Helmholtz coil to supply the dc bias magnetic field for the detection processes of both low-frequency magnetic and electric signals. An interesting advantage is the elimination of internal electrodes within the cell; further features will be discussed later.

We first demonstrate the proof-of-principle operation of the proposed method by sensing low-frequency magnetic signal with Rydberg level 60$^2$S$_{1/2}$, where $\Omega_c$ is about $2\pi\times 2.1~\mathrm{MHz}$ at the center of 480 nm beam. A separate coil feeds the low-frequency magnetic signals to the system for detection. Here, the dc bias magnetic field lifts the degeneracy of the $m_S=\pm1/2$ sublevels and puts the Zeeman shift into its linear regime for the incoming-signal-induced perturbations. The probe laser is right circularly polarized and the coupling laser is left circularly polarized. In Fig. \ref{fig:low_frequency_magnetic} we present a set of sample results. Fig. \ref{fig:low_frequency_magnetic}(a) shows the a typical trace of data where the dc bias magnetic field is 1.9 Gs and the exerted sinusoidal signal has amplitude of 8.3 Gs and frequency of 3.0 kHz. We obtain this spectrum by taking the Fast Fourier Transform (FFT) as the second layer of demodulation. This method does not require exact resonance of the incoming low-frequency signal with the Rydberg transitions and therefore we expect it to exhibit favorable characteristics in frequency dynamic range and instantaneous bandwidth. To check the actual performance on this point, we use the same apparatus to measure the frequency modulated signal and summarize the typical outcome in Fig. \ref{fig:low_frequency_magnetic}(b). In particular, we examine the response with respect to the linear-chirping signal and recovered signal on the time domain from demodulation.

\begin{figure}[h]
\centering
\includegraphics[width=0.46\textwidth]{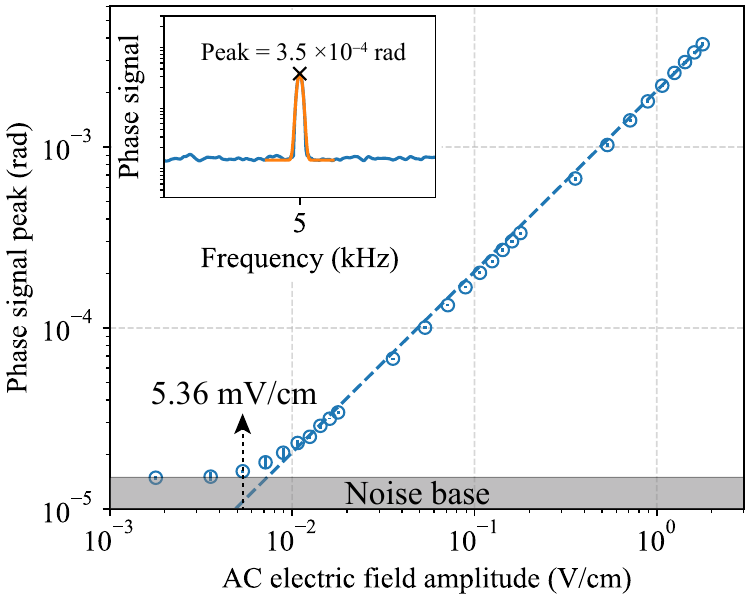}
\caption{(Color online) The spectrum peak strengths of demodulated phase signal with different electric signal amplitudes at frequency 5 kHz, under 18 Gs bias magnetic field. This series of measurement aims to evaluate the lowest detectable field and sensitivity. The error bars are one standard deviation from five independent experiments and the dashed line is linear fitting.}
\label{fig:electric_sensitivity}
\end{figure}

Now we discuss another main topic of sensing the low-frequency electric signal, which has vast opportunities in various applications. For improved sensitivity to electric fields and enhanced coupling strength with the driving laser, we choose to work with Rydberg D states as the destination of two-photon ground-Rydberg excitation. More specifically, we perform a series of experiments on the Rydberg state 58$^2$D${5/2}$ under a dc bias magnetic field and summarize the results in Fig. \ref{fig:electric_sensitivity}, where $\Omega_c$ is about $2\pi\times 3.7~\mathrm{MHz}$ at the center of 480 nm beam. The probe laser and the coupling laser are both vertically polarized. A set of metal field plates act as the source of low-frequency electric signal and the applied electric signal is fixed at 5 kHz, while its amplitude is varied to evaluate the sensitivity.

Our results contain one special feature of the ambient first-order response with respect to the incoming signal frequency. Namely, a low-frequency electric field with frequency $\omega_\text{l}$ will induce a phase modulation at frequency $\omega_\text{l}$. This may initially look surprising because a pure 58D state exhibits only a second-order Stark effect in weak fields. It turns out, according to our theoretical and experimental investigations, this first-order response essentially stems from the Rydberg dipole-dipole interactions that extensively exist in the atomic vapor. More specifically, we estimate that the average inter-atomic distance between two $^{87}$Rb is about 4 $\mu$m. At such distances or even closer, the Rydberg dipole-dipole interaction induces the F\"orster resonance between two atoms. Therefore, the destination state of the Rydberg EIT is no longer a purely 58D state but a linear superposition of D state plus nearby S, P, F states. Subsequently, the first-order Stark effect occurs which has linear dependence between the frequency shift and the electric field amplitude. This effect generates the first-order response in our heterodyne detection of low-frequency fields via Rydberg EIT with phase demodulation, where the first-order response is significantly stronger than the higher-orders as we observe experimentally.

The screening effect against external electric fields by the metallic alkali atom layer on the vapor cell interior shows a frequency-dependent behavior \cite{PhysRevApplied.13.054034} as we have also observed experimentally. Due to the screening effect, the low-frequency electric field experienced by atoms inside the glass cell constitutes only a small fraction of the external field. The lowest detectable signal strength and sensitivity with respect to purely the atom-field interaction needs to include the quantitative analysis of screening effect. We evaluate this via two experimental methods including utilization of the second-order Stark effect near the Rydberg EIT transparency point and calibration between the Rydberg EIT amplitude and phase signals. We calibrate the screening ratio to approximately 3.5\% at 5 kHz. 

We measure the spectrum peak strengths of demodulated phase signal as a function of incoming electric signal amplitude, where the peak value is read out from the FFT spectrum with total integration time of 0.1 s, as shown in Fig. \ref{fig:electric_sensitivity}. At this moment, the minimal detectable field amplitude is $5.36~\mathrm{mV/cm}$ outside the vapor cell. By calculating the SNR of the FFT spectrum, we get the sensitivity of our measurement to be $1.21~\mathrm{mV~cm^{-1}Hz^{-0.5}}$. With the above value of screening ratio, our result represents a minimal detectable electric field of $186~\mathrm{\mu V/cm}$ and sensitivity down to $42.4~\mathrm{\mu V~cm^{-1}Hz^{-0.5}}$ in-cell as seen by the $^{87}$Rb atoms.

\begin{figure}[h]
\centering
\includegraphics[width=0.46\textwidth]{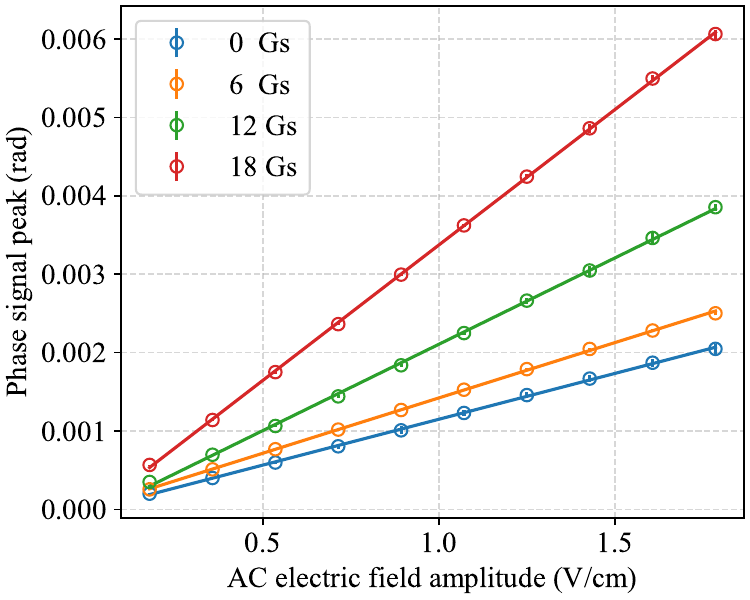}
\caption{(Color online) Performance with respect to different amplitudes for incoming electric signals at 5.0 kHz. The DC bias magnetic field is parallel to the AC electric field.
The error bars are one standard deviation from five independent experiments and the solid lines are linear fitting.
The slope from 0 Gs to 18 Gs is $(1.17 \pm 0.01)\times 10^{-3}, (1.41 \pm 0.01)\times 10^{-3}, (2.20 \pm 0.02)\times 10^{-3}, (3.45 \pm 0.01)\times 10^{-3}~\mathrm{rad~cm/V}$.}
\label{fig:electric_vary_bias}
\end{figure}

Here, applying the dc bias magnetic field also helps to lift the degeneracies and technically can even help to improve performance if the Zeeman-shifted states has better response to the electric field. In particular, we observe that even at 0 bias magnetic field the linear response hold well and such that it is not the cause of the first-order response. We also observe that the linearity holds well between the spectrum peak strengths of demodulated phase signal and the amplitudes of exerted electric fields within the shown range. This stems from the combination of linearity of EIT response as in Eq. \eqref{eq:EIT_1st_order} and the linearity of the first-order Stark effect as in Fig. \ref{fig:electric_sensitivity}.

\begin{figure}[h]
\centering
\includegraphics[width=0.46\textwidth]{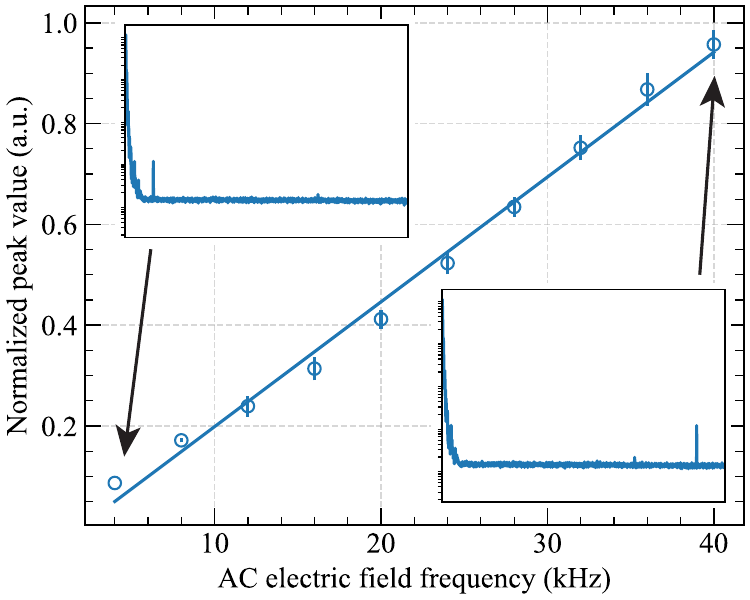}
\caption{(Color online) Performance with respect to different frequencies for incoming electric signals at a constant amplitude. The error bars are one standard deviation from five independent experiments and the solid line is to guide the eye.
The insets show the complete spectrum of demodulated phase signals.}
\label{fig:electric_frequency_scan}
\end{figure}

As stated previously, we expect this method to have a relatively large frequency response range, and we need to demonstrate this point experimentally. More specifically, we test the system with electric signals of different frequencies and obtain the result as shown in Fig. \ref{fig:electric_frequency_scan}. These results demonstrate compatibility with low-frequency electric signals in this initial experimental implementation. Moreover, the response of system is endowed with a clear frequency-dependent feature. One major cause is the frequency-dependence of the screening effect, namely that the metallic layer of the vapor cell functions like a high-pass filter for the incoming low-frequency electric signal \cite{Lim2023APL}. There also exist other factors that can possibly contribute to the frequency-dependence such as the demodulation instruments' intrinsic properties. Overall, we think that the main reason of seemingly linear dependence on frequency in Fig. \ref{fig:electric_frequency_scan} corroborates the behavior of screening effect. Namely, the frequency range under investigation in Fig. \ref{fig:electric_frequency_scan} is in fact relatively small compared with the full range where the screening effect has significant influences.

Currently, our results remain significantly distant from the standard quantum limit (SQL) of the system. While the screening effect currently constitutes the major challenge for low-frequency electric field detection, typical adverse effects associated with quantum sensing in atomic vapor, such as Doppler effects from different atomic velocity groups and inter-atomic collisions, can also influence our experiment. For example, the Doppler effects associated with different velocity groups of atoms and the inter-atomic collisions. Furthermore, realistic atomic states and linkage patterns are significantly more complex than those in the idealized three-level atom model. With these in mind, we will endeavor to experimentally establish a better Rydberg EIT within the system in the future.

In conclusion, we have designed and demonstrated an effective method for quantum sensing of low-frequency fields with Rydberg atoms via phase modulation and demodulation. It applies to both the low-frequency magnetic and electric fields. Our results hint at the possibility of integrating the detection of both low-frequency magnetic and electric fields into a single compact atom-based device. The information of incoming low-frequency fields is received as phase modulation of probe laser in the Rydberg EIT transparency window and the scalability of our method can become another important factor in potential applications. The observed linear response to the low frequency electric field signals is an example of how the many-body interactions help to enhance the performance of quantum sensing processes. At this moment, the measurement relies on the Rydberg dipole-dipole interactions of solely $^{87}$Rb. Extending this study to two- and multi-species systems \cite{PhysRevLett.119.160502, Yuan2024SCPMA} appears promising for exploring further improvements and additional functionalities.

\section*{Acknowledgments}
\label{sec:acknowledgments}

The authors gratefully acknowledge funding supports from the Science and Technology Commission of Shanghai Municipality (Grant No. 24DP2600202), the National Key R\&D Program of China (Grant No. 2024YFB4504002), Industrial Technology Development Research Program of Shanghai Institute of Optics and Fine Mechanics, and the National Natural Science Foundation of China (Grant No. 92165107). Shenchao Jin acknowledges the support from the China Postdoctoral Science Foundation under Grant No. 2024M753359 and Xin Wang acknowledges the support from the China Postdoctoral Science Foundation under Grant No. 2022M723270.

\bibliographystyle{apsrev4-2}

\renewcommand{\baselinestretch}{1}
\normalsize

\bibliography{trombone_ref}
\end{document}